\documentclass[%
 aip,
 amsmath,amssymb,
 reprint,%
]{revtex4-1}

\usepackage[utf8]{inputenc}
\usepackage[T1]{fontenc}
\usepackage{mathptmx}

\usepackage{mathrsfs}
\usepackage{amsmath,gensymb}
\usepackage{amsfonts}
\usepackage{amssymb}
\usepackage{amsthm}
\usepackage{graphicx}
\usepackage{natbib}
\usepackage{color}
\usepackage{hyperref}
\usepackage{bm}
\usepackage{verbatim}
\usepackage{dcolumn,subfig,mathtools}
\usepackage{soul}

\begin{document}

\title{Magnon Thermal Edelstein Effect Detected by Inverse Spin Hall Effect}
\author{Hantao Zhang}
\affiliation{Department of Electrical and Computer Engineering, University of California, Riverside, California 92521, USA}
\author{Ran Cheng}
\email{rancheng@ucr.edu}
\affiliation{Department of Electrical and Computer Engineering, University of California, Riverside, California 92521, USA}
\affiliation{Department of Physics and Astronomy, University of California, Riverside, California 92521, USA}

\begin{abstract}
In an easy-plane antiferromagnet with the Dzyaloshinskii-Moriya interaction (DMI), magnons are subject to an effective spin-momentum locking. An in-plane temperature gradient can generate interfacial accumulation of magnons with a specified polarization, realizing the magnon thermal Edelstein effect. We theoretically investigate the injection and detection of this thermally-driven spin polarization in an adjacent heavy metal with strong spin Hall effect. We find that the inverse spin Hall voltage depends monotonically on both temperature and the DMI but non-monotonically on the hard-axis anisotropy. Counterintuitively, the magnon thermal Edelstein effect is an even function of a magnetic field applied along the Néel vector. 
\end{abstract}

\maketitle

Efficient generation of spin angular momenta is key to modern electronics because information encoded in the spin degree of freedom can be transported and read with relatively low dissipation~\cite{spintronics_review, interface_magnetism_review}. In electronic systems, there are two common scenarios to produce spins. One is the spin Hall effect (SHE), in which a charge current induces a pure spin current flowing transversely; as spin current diffuses, surviving spins will accumulate on the boundaries~\cite{spin_hall_review}. The other is the Edelstein effect, where spin accumulation can be generated directly from a charge current without the participation of spin current~\cite{Edelstein_effect, SEE_2, SEE_NatPhy, ISEE_1, ISEE_2}. Although the spin accumulation may experience diffusion, at the stage of generation the Edelstein effect does not entail spin diffusion in space. In large-scale devices, spin diffusion can be a prevailing cause of information loss~\cite{Maekawa}. Therefore, concerning spin production, the Edelstein effect appears to be more reliable than the SHE.

Nevertheless, both the SHE and the Edelstein effect inevitably incur Joule heating because a driving charge current is needed. It has been established that magnons, which are charge neutral quasiparticles in magnet materials, can conduct spins without physical movement of charges, eliminating Joule heating completely~\cite{magnon_spintronics_review,Kajiwara,Kruglyk}. Consequently, magnons are believed to be an ideal alternative to electrons in shaping the next-generation nanodevices such as all-magnon transistor~\cite{Chumak}. A recent highlight in the quest for magnon-based spin transport is the identification of antiferromagnets (AFMs) as a platform superior to ferromagnets owing to their unique characteristics such as ultrafast spin dynamics and absence of stray field~\cite{AFM_spintronics_review, Review_Rezende}. In particular, AFMs allow the coexistence of two distinct spin species constituting an intrinsic degree of freedom, which is capable of encoding information similar to the electron spin~\cite{AFM_magnon_1, magnon_transistor}.

A magnon spin current can be either coherent or incoherent depending on the preservation of transverse spin polarization. In AFMs, driving coherent magnons calls for extremely high frequency sources~\cite{AFM_pumping1,AFM_pumping2}, whereas incoherent magnons can be easily generated through thermal agitations~\cite{Review_Rezende}. For example, in magnetic insulators exhibiting the magnon spin Nernst effect, a temperature gradient can create a pure spin current in the perpendicular direction, which is analogous to the SHE of electrons~\cite{magnon_spin_nernst_1, magnon_spin_nernst_2, MnPS3}. So far, however, magnon-based spin generations all rely on the formation of spin currents~\cite{MnPS3,SSE1,SSE2,SSE3}. It is unclear whether non-equilibrium spin accumulation can arise directly from a thermal drive free of an accompanied spin current. Namely, is there a magnonic counterpart of the Edelstein effect, or magnon thermal Edelstein effect (TEE)?

\begin{figure}[t]
	\centering
  		\includegraphics[width=\linewidth]{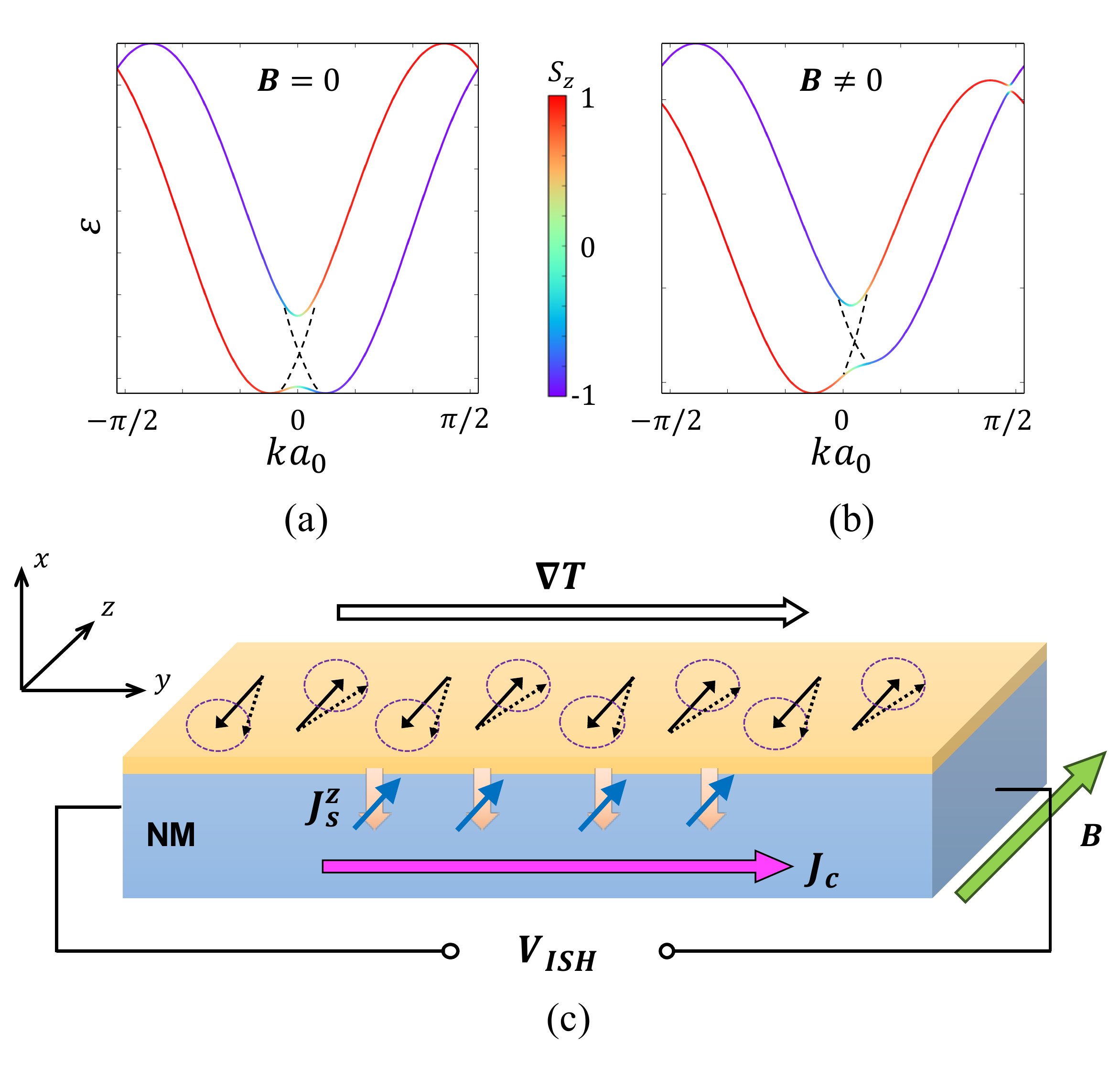}
	\caption{An illustration of the dispersion relations for (a) $B=0$ and (b) $B\neq0$. The dotted segments indicate the case of $K_x=0$. $K_x$ and $D$ are exaggerate for visual clarity. The color labels $S_z$ as a function of $k$. (c) Schematics of the model 1D AFM/NM heterostructure. While the spin current along $\nabla T$ vanishes in the AFM, the spin accumulation $S_z$ converts into a spin current in the NM flowing in the $x$ direction.}
	\label{fig:band_model}
\end{figure}

Recently, it has been pointed out that the magnon TEE can indeed be enabled by breaking certain symmetries in a collinear AFM~\cite{magnon_Edelstein}. Specifically, in the presence of the Dzyaloshinskii-Moriya (DM) interaction, magnon bands of opposite spins shift oppositely in the momentum space, leading to the Faraday effect of spin waves~\cite{magnon_transistor}. On top of the DM interaction, if non-uniaxial magnetic anisotropy is also introduced, the magnon spectrum will be similar to that of the gaped Rashba electrons~\cite{electron_SOC, magnon_Rashba} [see Fig.~\eqref{fig:band_model}(a)]. The combined effect of the DM interaction and the non-uniaxial anisotropy can lock spins with momenta, which is critical to the TEE. While a magnon spin current can convert into an electron spin current and be subsequently detected via the inverse SHE, it remains an open question how to measure the magnon TEE that is not accompanied by a spin current.

In this Letter, we consider an ultrathin AFM nanostrip in contact with a normal metal (NM) with a sizable SHE, as schematically illustrated in Fig.~\eqref{fig:band_model}(c). Because of the magnon TEE, an in-plane temperature gradient $\nabla T$ can drive the non-equilibrium magnon accumulation which will couple the electrons in the NM, while magnon motion perpendicular to the interface is suppressed. It will become clear later that the thermal current induced by $\nabla T$ is not spin polarized, so the AFM admits only spin accumulation but not spin current. This magnon accumulation injects a pure spin current in the NM, which, due to the inverse SHE, will convert into a detectable voltage $V_{ISH}$. We find that the magnon TEE depends monotonically on temperature and the DM interaction but non-monotonically on the hard-axis anisotropy. Counterintuitively, the magnon TEE turns out to be an even function of an applied magnetic field along the N\'{e}el vector, which is followed by an explanation from symmetry perspective.

The AFM nanostrip under consideration is modeled as an infinitely long 1D spin chain along $y$ direction with atomic spacing $a_0$ (so magnetic unit cell repeats for $2a_0$). As illustrated in Fig.~\ref{fig:band_model}(c), spins in the ground state are pointing in an alternative pattern along the $z$ direction, which can be stabilized by an easy-axis along $z$ and a hard-axis along $x$. The broken mirror symmetry on the interface induces a DM interaction with the DM vector parallel to $z$, the strength of which is partially controllable by a gate voltage~\cite{Nagaosa,E_field_DMI,DMIbyE}. In the following, we assume that $\nabla T$ is along $y$ and a magnetic field $\bm{B}$ is in the $z$ direction. By keeping only the nearest-neighboring interactions, we can describe such an AFM by a tight-binding Hamiltonian as
\begin{align} \label{Hamiltonian_H}
H =& J\sum\limits_{<i,j>} \bm{S}_{i} \cdot \bm{S}_{j} + \sum\limits_{i}  \left[K_xS_{i,x}^2-K_zS_{i,z}^2\right] \notag\\
&+ D\sum\limits_{<i,j>} \xi_{ij} \hat{\bm{z}}\cdot \left( \bm{S}_{i} \times \bm{S}_{j} \right) - B \sum\limits_{i} \hat{\bm{z}}\cdot\bm{S}_i,
\end{align}  
where $J$ is the antiferromagnetic exchange coupling, $D$ is the DM interaction, and $K_x$ and $K_z$ are the hard-axis and the easy-axis anisotropy, respectively. In the DM term, $\xi_{i,i\pm1}=\pm1$ represents the relative positions of neighboring spins (bond chirality). For simplicity, we have absorbed the Land\'e $g$ factor and the Bohr magneton $\mu_{B}$ into $B$ so that all parameters appearing in Eq.~\eqref{Hamiltonian_H} are scaled into the energy dimension.

To solve the magnon excitations, we perform the linearized Holstein-Primakoff transformation on the spin operators~\cite{HP_transform}: $S^{+}_{iA} = \sqrt{2S} a_{i}$, $S^{-}_{iA} = \sqrt{2S} a^{\dagger}_{i}$, $S^{z}_{iA} = S - a^{\dagger}_{i} a_{i}$, $S^{+}_{i+1 B} = \sqrt{2S} b^{\dagger}_{i+1}$, $S^{-}_{i+1 B} = \sqrt{2S} b_{i+1}$, and $S^{z}_{i+1B} = -S + b^{\dagger}_{i+1} b_{i+1}$, where $a_i^\dagger$ ($b_i^\dagger$) creates a quanta of magnon on the $A$ ($B$) sublattice located on site $i$. The magnon operators satisfy the bosonic commutation relations: $[a_i,a_j^\dagger]=\delta_{ij}$; and all other combinations vanish.

To diagonalize Eq.~\eqref{Hamiltonian_H}, we perform the Fourier transformation $a_{i} = \frac{1}{\sqrt{N}} \sum_{k} a_{k} e^{iky_{i}}$, $b_{j} = \frac{1}{\sqrt{N}} \sum_{k} b_{k} e^{iky_{j}}$ where $N$ is the total number of magnetic unit cells, $y_{i}$ labels the position of the $i$th atom, and the summation over $k$ is restricted to the first Brillouin zone. The Hamiltonian in Eq.~\eqref{Hamiltonian_H} can be expressed in a quadratic form
\begin{align} \label{Hk}
H = \frac12 \sum_{k} X_k^\dagger
\begin{pmatrix*}[l]
A^+ & F & 0 & C_{k} \\
F & A^+ & D_{k} & 0 \\
0 & D_{k} & A^- & F \\
C_{k} & 0 & F & A^-
\end{pmatrix*} 
X_k,
\end{align}
where $X_k=[a_k,a_{-k}^\dagger,b_k,b_{-k}^\dagger]^T$ is the Nambu basis, $A^{\pm} = 4JS + 2K_{z}S + K_{x}S \pm B$, $F = K_{x}S$, $C_{k} = 4JS\cos k a_{0} + 4DS \sin k a_{0}$ and $D_{k} = 4JS\cos k a_{0} - 4DS \sin k a_{0}$. Because of the commutation relations, the vector of the Nambu basis must satisfy 
\begin{align} \label{commutation}
\left[ X, X^{\dagger} \right] = 
\begin{pmatrix*}[c]
1 & 0 & 0 & 0 \\
0 & -1 & 0 & 0 \\
0 & 0 & 1 & 0 \\
0 & 0 & 0 & -1 
\end{pmatrix*} \equiv g_c,
\end{align}
where the $(i,j)$-component of $[X,X^\dagger]$ reads $[X_i,X_j^\dagger]$. Diagonalizing the quadratic form in Eq.~\eqref{Hk} invokes the Bogoliubov transformation, which amounts to a coordinate change of the Nambu basis~\cite{Review_Rezende}: $\tilde{X}_k = Q_{k}X_k = [\alpha_k,\alpha_{-k}^\dagger,\beta_k, \beta_{-k}^\dagger]^T$, where $Q_k$ satisfies
\begin{align} \label{intro_Qk}
Q_k^{-1}g_c H_k Q_k = \hbar g_c \mathrm{diag}[\omega_k^\alpha,\omega_{-k}^\alpha,\omega_k^\beta,\omega_{-k}^\beta],
\end{align}
where $\omega_{\pm k}^{\alpha/\beta}$ are the eigenfrequencies (dispersion relations) and $Q_{k}$ is a generalized unitary matrix with respect to $g_c$: $Q_{k} g_c Q^{\dagger}_{k} = g_c$. Under the same Fourier transform, the $z$ component of the total spin operator becomes 
\begin{align} \label{spin_z}
    S_{z} =\sum_iS_{i,z} = \frac{1}{2} \sum_{k} X^{\dagger}_{k} \sigma_{z} X_{k},
\end{align}
where $\sigma_z=\mathrm{diag}[-1,-1,1,1]$.

Figure~\ref{fig:band_model}(a) and (b) illustrate the dispersion relations (parameters exaggerated to enhance visual clarity), where the magnon spin polarization $S_z(k)$ is indicated by color. If not for the hard axis, the $\alpha$ and $\beta$ bands will have constant $S_z$ ($k$-independent) and intersect at $k=0$, as indicated by the dotted line; their band bottoms locate on the opposite side of $k=0$ due to the DM interaction. The hard-axis anisotropy $K_x$ opens a gap at $k=0$, which re-organizes the bands into a lower branch ($\beta$) and an upper branch ($\alpha$). Correspondingly, $S_z(k)$ becomes a function of $k$ in the vicinity of the band gap where it transitions continuously between $1$ and $-1$. Within the energy window of the band gap at $k=0$, magnons with spin $S_z=1$ ($S_z=-1$) are locked with a negative (positive) group velocity $d\omega/dk$ in the outer segments of the lower branch. In contrast, above the band gap, the two branches contribute oppositely to $S_z$ so that both $S_z=1$ and $S_z=-1$ can be associated with a given $d\omega/dk$. We will see that this spin-momentum locking is essential to the magnon TEE.

In an AFM/NM heterostructure, coherently precessing magnetic moments inject a pure spin current density in the NM~\cite{Ran_pumping,Brataas_pumping} 
\begin{equation} \label{eq:sp}
\bm{J_{s}} = g_s \left( \bm{n} \times \dot{\bm{n}} + \bm{m} \times \dot{\bm{m}} \right),
\end{equation}
where $\bm{n} = ( \bm{S}_{A} - \bm{S}_{B} )/2$ and $\bm{m} = ( \bm{S}_{A} + \bm{S}_{B} )/2$ are the N\'{e}el vector and the small magnetization, $g_s$ is the real part of the spin-mixing conductance (factors like $\hbar$, $e$ have been absorbed by $g_s$). Even though Eq.~\eqref{eq:sp} was derived for spatially uniform AFM, it remains valid even when the order parameter varies smoothly in space~\cite{uniform_pumping}. Because the TEE only generates incoherent thermal magnons, we need to insert the Holstein-Primakoff transformation into Eq.~\eqref{eq:sp} and integrate over all magnon modes~\cite{Rezende_pumping}. The key distinction of incoherent magnons is that the local spin component transverse to the easy-axis is averaged to zero so that only the $z$ component matters. After some tedious algebra, we obtain the spin current density injected into the NM as
\begin{align} \label{current_thermal}
J_{s}^{z} =& -\frac{g_{s}S}{2 \hbar} \sum_{k} \mathrm{Tr} \{  (Q_k^{\dagger}J_k Q_k) \,  \mathrm{diag}[\langle\alpha_k^\dagger\alpha_k\rangle, \langle\alpha_{-k}^\dagger\alpha_{-k}\rangle,\notag\\ 
&\qquad\qquad\langle\beta_k^\dagger\beta_k\rangle, \langle\beta_{-k}^\dagger\beta_{-k}\rangle] \}
\end{align}
with
\begin{align}
J_k=
\begin{pmatrix*}[l]
A^{+} & F & 0 & 0 \\
F & A^{+} & 0 & 0 \\
0 & 0 & -A^{-} & -F \\
0 & 0 & -F & -A^{-}
\end{pmatrix*},
\end{align}
where $\langle \alpha_{k}^{\dagger} \alpha_{k} \rangle=n(\hbar \omega^{\alpha}_{k})$ and $\langle \beta_{k}^{\dagger} \beta_{k} \rangle=n(\hbar \omega^{\beta}_{k})$ are the thermal occupations of magnons in the presence of $\nabla T$. To characterize the efficiency of spin injection into the NM, we define the spin convertance as
\begin{align} \label{eq:Gs}
    G_s=J_s^z/S_z
\end{align}
which directly converts the spin accumulation generated by $\nabla T$ into an electronic spin current ready for detection. This quantity plays a central role in quantifying the detection of the magnon TEE.

In the linear response regime, we can decompose the distribution as $n = n^0+\delta n$ with the equilibrium distribution $n^{0} = 1/\left(e^{\hbar \omega / k_{B}T} -1\right)$. Under the relaxation time approximation, the non-equilibrium part is
\begin{equation} \label{non_eq_dis}
\delta n_{\lambda,k} = -\tau_0 v_{\lambda, k} \frac{e^{\hbar \omega_{\lambda, k} / k_{B}T} \hbar \omega_{\lambda, k}}{k_{B} \left( e^{\hbar \omega_{\lambda, k} / k_{B}T} -1 \right)^{2} T^{2}} \nabla T,
\end{equation}
where $\lambda = \alpha$ or $\beta$ label the magnon modes, $T$ is the equilibrium temperature at the system center, $v_{\lambda, k}=\partial\omega_{\lambda}/\partial k$ is the group velocity of mode $\lambda$, and $\tau_0$ is the phenomenological magnon relaxation time assumed to be an independent constant. Based on Eq.~\eqref{non_eq_dis}, it should be noted that the in-plane spin current along $\nabla T$ (spin Seebeck effect) identically vanishes because when integrating over the Brillouin zone, the integrand is proportional to  $v_{\lambda,k}^{2}$ so $\left<S_{z}\right>$ from $\pm k$ cancel~\cite{Review_Rezende, magnon_spin_seebeck}. In other words, $\nabla T$ can only induce a spin accumulation $S_z$ but not a spin current in the AFM without the magnetic field. It is worth mentioning that the broken mirror symmetry along $x$ yields the DM interaction $D\hat{\bm{z}}$ along the N\'{e}el order, which shifts the magnon bands along $k_{y}$ direction. Therefore, only a temperature gradient along $y$ will induce spin imbalance, which is reflected in Eq.~\eqref{non_eq_dis}.

\begin{figure}[t]
	\centering
  		\includegraphics[width=\linewidth]{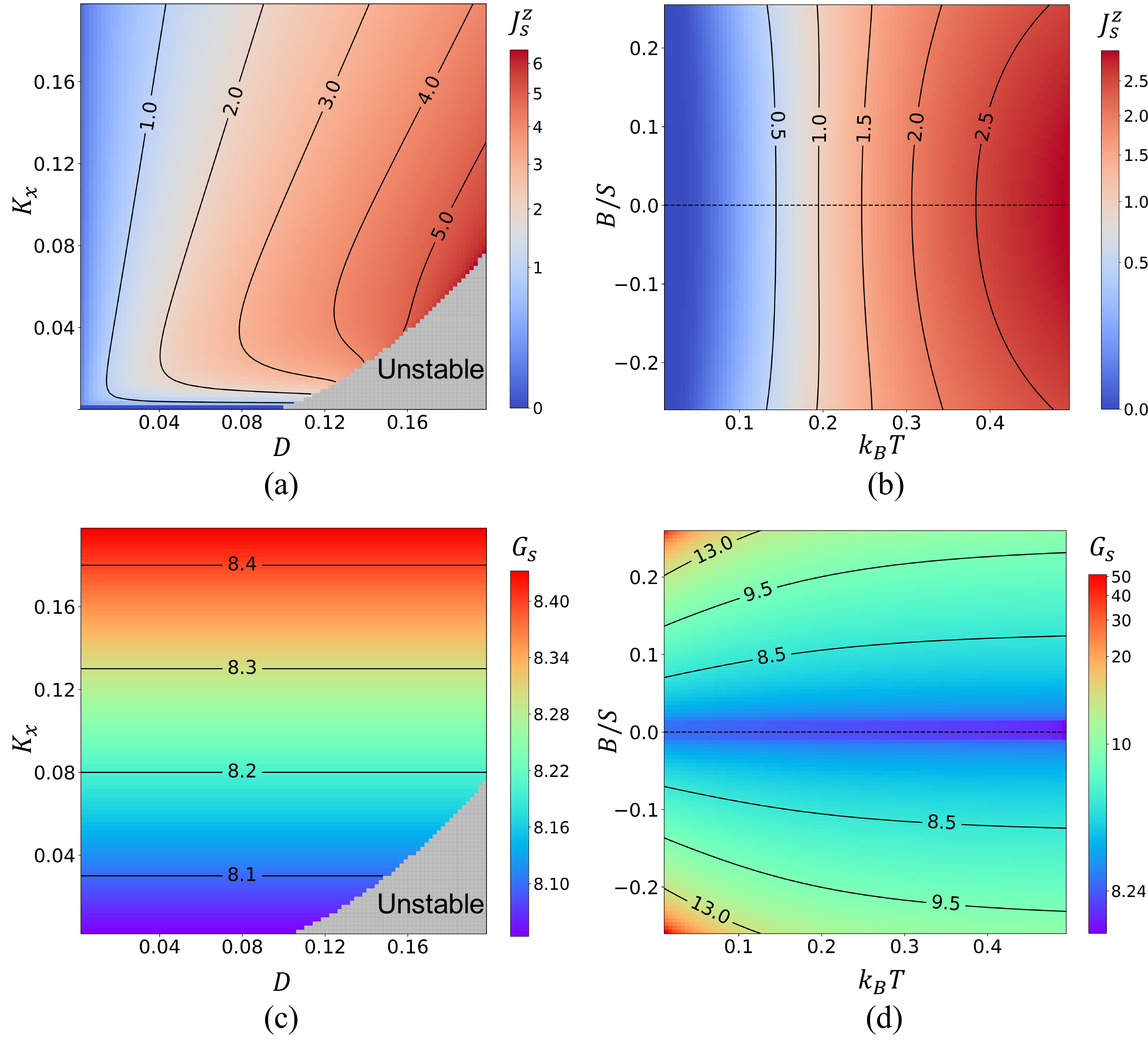}
	\caption{Spin current density injected into the NM, $J_s^z$, plotted (a) as a function of $D$ and $K_{x}$ for $k_{B}T=0.5$ and $B = 0$, and (b) as a function of $k_{B}T$ and $B/S$ for $K_{x} = 0.1$ and $D = 0.1$. (c) and (d): spin convertance $G_s$ plotted as functions of the same variables in (a) and (b). Units: $J_s^z$ in $g_{s}S^{4} \tau_{0} a_{0} k_{B} \nabla T / 2\hbar^{2}$ and $G_s$ in $g_{s}S^{2}/\hbar$. The grey regions are where the collinear ground state becomes unstable and the modeling becomes invalid.}
	\label{fig:D_Kx_T_dep}
\end{figure}

The spin accumulation $S_z$ due to the magnon TEE converts into a spin current into the NM according to Eq.~\eqref{eq:Gs}, which is subject to diffusion and the inverse SHE. By solving the spin diffusion equation with $J_s^z$ obtained from Eq.~\eqref{current_thermal} at the AFM/NM interface and vanishing spin current at the opposite interface of NM as the boundary conditions~\cite{boundary_Bauer_1, boundary_Bauer_2, boundary_Bauer_2013,Boundary}, we obtain the output inverse SHE voltage as
\begin{align}\label{ISHE_voltage}
    V_{ISH}=\frac{\theta_{SH}L\lambda}{\sigma d}\tanh\frac{d}{2\lambda}J_s^z,
\end{align}
where $\theta_{SH}$ is the spin Hall angle, $d$ is the thickness of the NM, $\lambda$ is the spin diffusion length, $\sigma$ is the conductivity, and $L$ is the length in the $y$ direction. We stress that Eq.~\eqref{ISHE_voltage} is a general solution of the spin diffusion equation regardless of the coherency of magnons. By inserting Eq.~\eqref{current_thermal} into Eq.~\eqref{ISHE_voltage}, we are able to quantify the output signal. Admittedly, treating $\tau_0$ as a quantity independent of momentum, spin, and temperature is an oversimplified assumption. Nevertheless, the modeling has captured the essential physical feature of the magnon TEE. Moreover, the overall output signal $V_{ISH}$ as a function of $T$ is determined by a series of quantities such as $\sigma$, $\lambda$ and $\theta_{SH}$, all of which have complicated $T$-dependence. Therefore, it goes beyond the scope of this Letter to give an exact $T$-dependence of $V_{ISH}$. Our focus is paid on the physical mechanism of the magnon TEE and its detection by the inverse SHE.

Figure~\ref{fig:D_Kx_T_dep}(a) plots the numerical result of $J_s^z$ as a function of the DM interaction and the hard-axis anisotropy for a fixed temperature. We find that the injected spin current in the NM depends monotonically on $D$ but non-monotonically on $K_x$. The $D$ dependence is easy to understand because $D$ shifts the magnon bands of opposite spins towards opposite directions in the momentum space, enhancing the spin-contrasting feature in the dispersion. The role of $K_x$, on the other hand, is complicated. As discussed previously, $K_x$ opens a gap at $k=0$ and gives rise to spin-momentum locking such that $S_z(k)$ is a function of $k$. For small $K_x$, an increasing gap widens the energy window of spin-momentum locking, which enlarges the spin imbalance hence a stronger magnon TEE. When $K_x$ becomes comparable to $D$, however, the transition region of $S_z(k)$ [$S_z(k)\approx0$ in the vicinity of $k=0$] will be extended to the outer part of the lower band, which reduces the magnitude of $S_z$ in the energy window of spin-momentum locking. Therefore, even though the spin-momentum locking still gets stronger with an increasing $K_x$, the net spin accumulation starts to decline. Using typical material parameters of transition metal oxides for the AFM and those of Pt for the NM, and assuming $g_{s} \sim 1/a^{2}_{0}$, $\tau_{0} \sim 10^{-10}\,\text{s}$, $d\gg\lambda$, and $L\sim1\mu m$, we estimate that a temperature gradient of $\nabla T \sim 1\text{K}/\mu m$ can generate a $V_{ISH}$ on the order of $100\,\text{nV}$ at room temperature, which is amenable to measurement.

\begin{figure}[t]
	\centering
		\includegraphics[width=\linewidth]{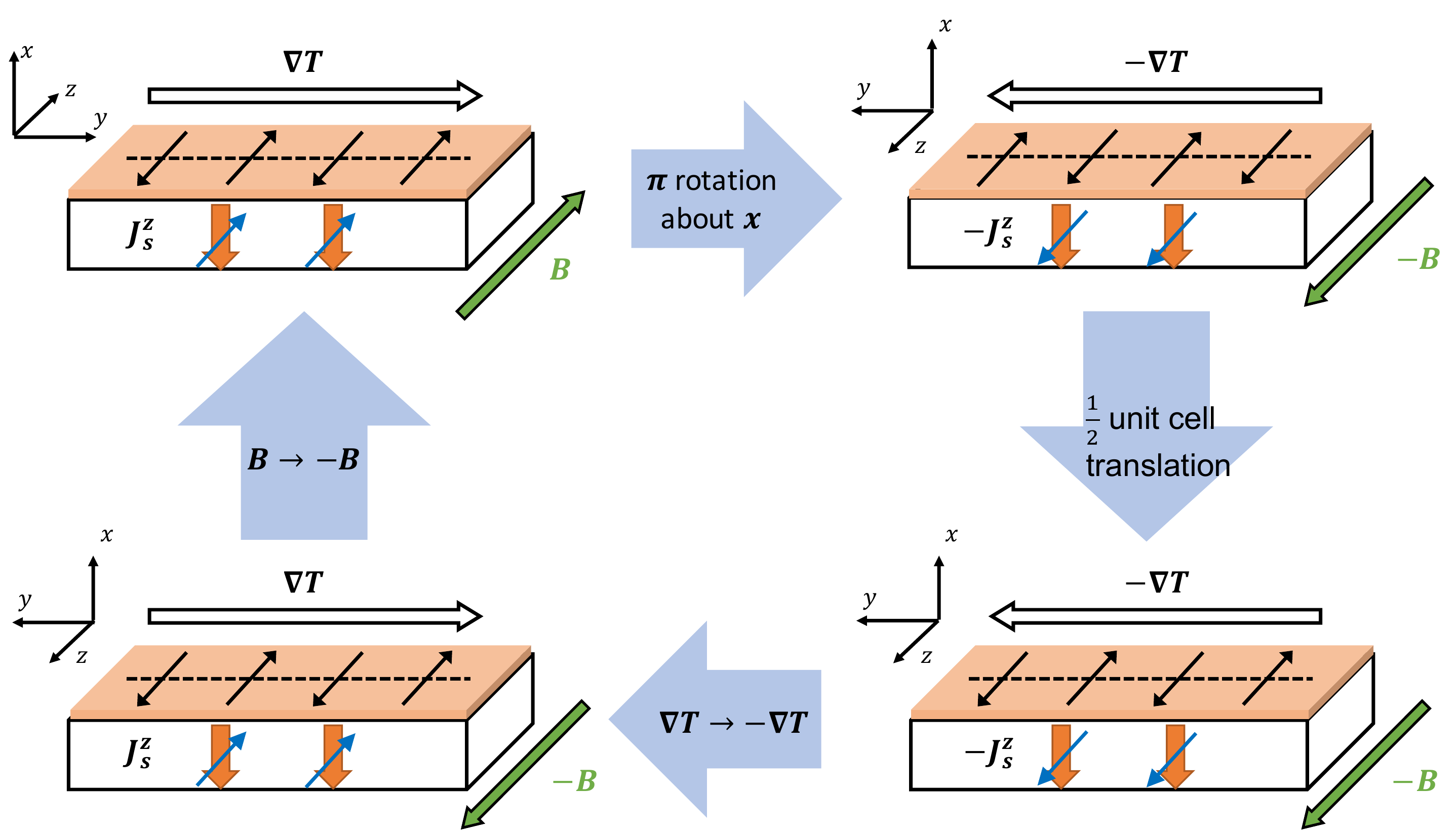}
	\caption{Symmetry analysis for the understanding of the magnon TEE as an even function of the magnetic field.
	 }
	\label{fig:B_field_symmetry}
\end{figure}

As shown in Fig.~\ref{fig:D_Kx_T_dep}(b), $J_s^z$ increases monotonically with an increasing temperature and is an even function of the applied magnetic field. Even concerning the bosonic statistics of magnons, this positive temperature dependence is not a trivial property, because comparatively the magnon spin Nernst effect~\cite{magnon_spin_nernst_1,magnon_spin_nernst_1,MnPS3} exhibits a non-monotonic temperature dependence. Before explaining the dependence of the magnon TEE on the magnetic field, it is instrumental to quantify the spin convertance $G_s$.

Figure~\ref{fig:D_Kx_T_dep}(c) and (d) plot $G_s$ versus the same variables used in Fig.~\ref{fig:D_Kx_T_dep}(a) and (b). According to Eq.~\eqref{eq:Gs}, $G_s$ is what connects the spin accumulation $S_z$ to the spin current density it induces in the NM. To quantify the magnon TEE, it is equally legitimate to use either $S_z$ or $J_s^z$. If plotted, the $S_z$ profile would be equivalent to (a)/(c) and (b)/(d). Similar to Fig.~\ref{fig:D_Kx_T_dep}(b), $G_s$ is also an even function of the magnetic field.

Now we explain the counterintuitive impact of the magnetic field shown in Fig.~\ref{fig:D_Kx_T_dep} from a symmetry perspective as illustrated in Fig.~\ref{fig:B_field_symmetry}. Starting from the upper-left panel, suppose that the injected spin current density is $J_s^z(B)$ for $B>0$ along $+z$. Our goal is to check if $J_s^z(-B)$ equals $J_s^z(B)$ or not. The cases of $B$ and $-B$ can be connected by a series of symmetry operations. First, we rotate the whole system around the $x$ axis by $\pi$, which flips the sign of everything, including $J_s^z$. Next, we shift the AFM chain by $a_0$ (half magnetic unit cell), which amounts to flipping the N\'eel vector if the system is infinite. This operation does not change $J_s^z$. Then, we reverse the direction of $\nabla T$, which reverses $J_s^z$ within the linear response regime. At this point, $J_s^z$ has flipped twice while the magnetic field $B$ has flipped once; the system only differs from where we started in the magnetic field direction. In other words, $J_s^z(B)=J_s^z(-B)$, the magnon TEE is even in the magnetic field.

In summary, we have studied the magnon TEE in a simple model system. An in-plane temperature gradient can drive magnon spin accumulation without inducing a magnon spin current. The spin accumulation converts into an electronic spin current in an adjacent NM for detection by the inverse SHE. The magnonic TEE depends monotonically on temperature and the DM interaction but non-monotonically on the hard-axis anisotropy. The effect is even in a magnetic field along the N\'{e}el order, which bears a symmetry justification. Our prediction opens the exciting possibility to generate magnonic spin angular momenta without inducing magnonic spin currents.

This work is supported by the Air Force Office of Scientific Research under grant FA9550-19-1-0307, and in part by the University of California, Riverside. Data sharing is not applicable to this article as no new data were created or analyzed in this study.

\end{document}